\documentclass[preprint,preprintnumbers,amsfonts,amsmath,amssymb,prd
,tightenlines]{revtex4}
\usepackage{graphicx}
\begin{document}
\preprint{FAU-TP3-08-06}
\title{Gauge fields in accelerated frames$^{{\rm * }}$}
\author{F. Lenz $^{{\rm **}}$}    
\affiliation{Institute for Theoretical Physics III \\
University of Erlangen-N\"urnberg \\
Staudstrasse 7, 91058 Erlangen, Germany\\}
\date{August 20, 2008}
\begin{abstract}
Quantized fields in accelerated frames (Rindler spaces) with emphasis on gauge fields are investigated. Important properties of the dynamics  in Rindler spaces are shown  to follow from the scale invariance of the corresponding Hamiltonians. Origin and  consequences of this extraordinary property of Hamiltonians in Rindler spaces are elucidated. Characteristics of the Unruh radiation, the appearance of a photon condensate and the interaction energy of vector and scalar static charges are discussed and implications for Yang-Mills theories and QCD in Rindler spaces are indicated. 
\vskip 11cm \hspace{-.2cm}\small{*\, To appear in the the proceedings of CAQCD08} \\ \small{** \,flenz@theorie3.physik.uni-erlangen.de}
\end{abstract} 
\maketitle
\section{Introduction} 
I will report   on an investigation   of gauge fields  in static space times   which has been carried out  in collaboration with K. Yazaki and K. Ohta \cite{LEOY08}. Our interest in this subject was triggered by the Ads/CFT correspondence which opened the possibility to formulate effective theories of e.g. QCD  in terms of quantum fields in gravitational backgrounds. The studies I report on are intended to improve and extend our understanding of the peculiar properties of the dynamics of quantized fields in static space-times.  

The following discussion will   focus on results concerning quantized gauge fields in Rindler spaces or equivalently  quantized gauge fields  as seen by a uniformly accelerated observer. The metric of Rindler spaces  incorporates the physics of the relativistic generalization of a homogeneous gravitational field and can therefore be used locally as an approximation to more complicated spaces such as the space close to the horizon of a Schwarzschild black hole. Since the discovery of the so called Unruh effect \cite{FULL73, DAVI75,UNRU76},  i.e. the appearance of a heat bath as a result of the acceleration and its relation to Hawking radiation  about 30 years ago, conceptual questions concerning the presence or absence of radiation have played an important role (for a review cf. \cite{CRHM07}).  Applications of such studies have addressed the motion of particles in accelerators  (cf.\,\cite{BELE87}), acceleration induced decays of particles \cite{MULL97,VAMA00} or the possibility of thermalization in relativistic heavy ion collisions \cite{KATU05}. While  most of the results  have been obtained  for scalar fields  no systematic investigations of the properties of gauge fields in Rindler spaces have been carried out. 
\section{Kinematics}
\begin{figure}
\includegraphics[width=.5\linewidth]{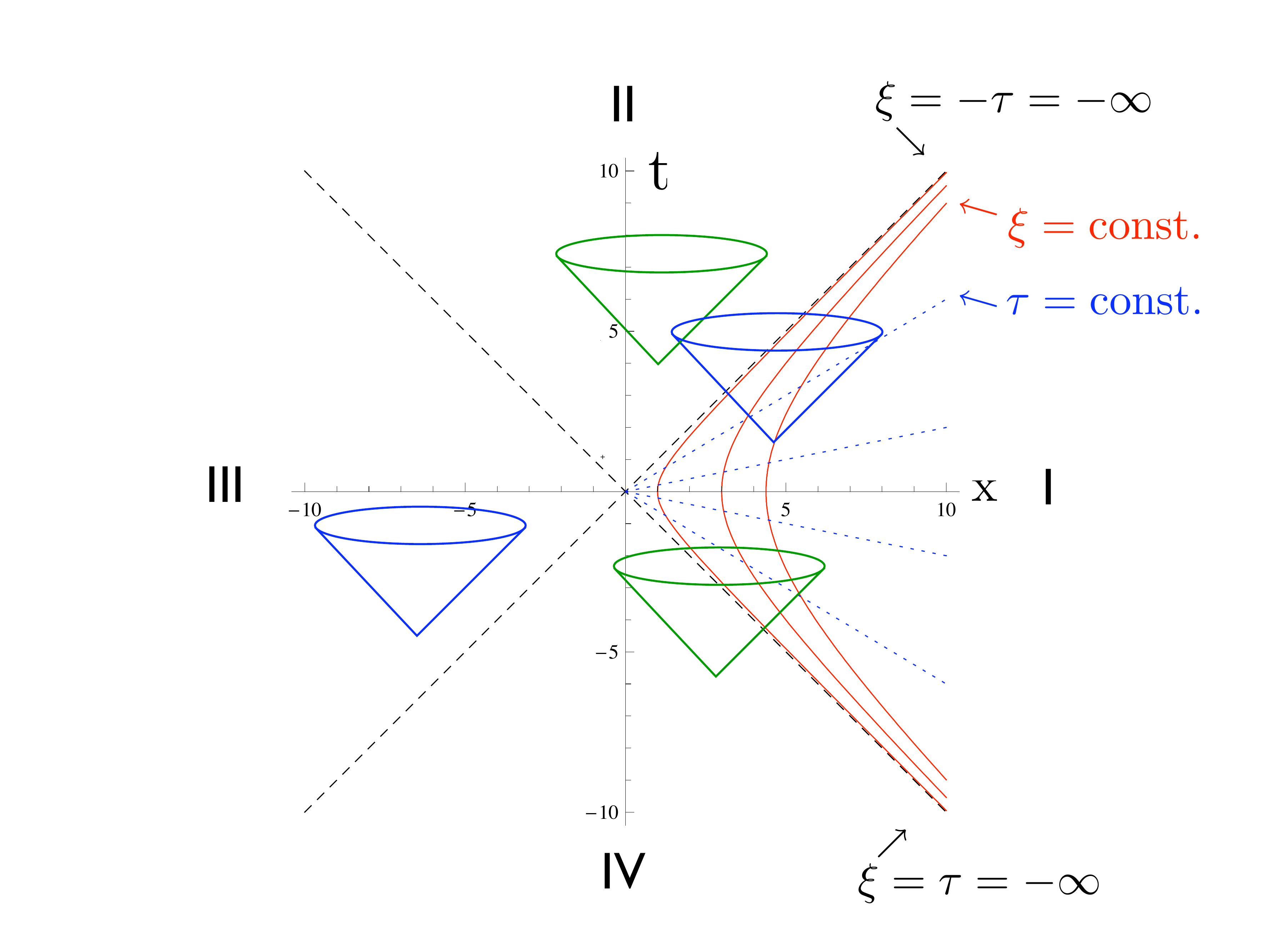}
\caption{Kinematics of uniform acceleration}
\label{kin}
\end{figure}

A uniformly accelerated observer in $d+1$ dimensional  Minkowski space moves along the hyperbola \cite{RIND01}
\begin{equation}x^{2} - t^{2} = \frac{1}{a^{2}}\,,\quad  
{\bf x}_{\perp}= 0\,. 
\label{hyper}
\end{equation}
The acceleration is denoted by $a$ and the d-1 coordinates  transverse to the motion by  ${\bf x}_{\perp}$ .
The initial conditions 
\begin{equation*} x(0)=\frac{1}{a}\,,\quad\frac{d x}{dt}\Big|_{t=0}=0\,,\end{equation*}
have been chosen.
To describe  quantum fields as seen by the accelerated observer we transform into his rest frame and consider  the coordinate transformation 
\begin{equation}t,x,x_{\perp}\to \tau,\xi,x_{\perp}:\quad t (\tau, \xi) = \frac{1}{a} \; e^{a \xi} \sinh a \tau\,, \quad 
x (\tau, \xi) = \frac{1}{a} \; e^{a \xi} \cosh a \tau \,,
\label{rimi}
\end{equation}
with the inverse
\begin{equation*}x^{2} - t^{2} = \frac{1}{a^{2}} \; e^{2 a \xi}\,,\quad
\frac{t}{x} = \tanh a\tau\,.\end{equation*}
By construction,  $\xi=0$ corresponds to the hyperbolic motion (\ref{hyper}) or, more generally, a particle at rest in the observers system at $\xi=\xi_0=$const.  corresponds to the uniformly accelerated motion  in Minkowski space with acceleration $a\exp\{-a\xi_0\}$.  Trajectories of uniformly accelerated particles for different values of $\xi_0$ are shown in Fig.\,\ref{kin} together with the lines $\tau=$const.\,.
We note that the mapping (\ref{rimi}) is not one-to-one. The coordinates $-\infty < \tau, \xi < \infty $ cover only one quarter of the Minkowski space, the right ''Rindler wedge''$R_+$ (region I)
\begin{equation}
R_{\pm} = \big\{x^\mu\big|\,|t|\le \pm x \big\}\, .                          
\label{rau3}
\end{equation}
Upon reversion of the sign of $x$ in Eq.\,(\ref{rimi})  the left Rindler wedge $R_-$ (region III) is covered by the corresponding parametrization. As illustrated in the Figure, the light cone $x=t $ corresponding to $\xi=-\tau=-\infty$ is  an event horizon. The light cones indicate that in region I  signals  can be transmitted  to  region II but not received from it. Signals received from region IV appear to have originated from the horizon $\xi=\tau= -\infty$.  
The space-time defined by the coordinate transformation (\ref{rimi}) is called Rindler space and its metric is given by
\begin{equation}ds^{2} = e^{2 a \xi} (d \tau^{2} - d \xi^{2})- d{\bf x}^2_{\perp}\,. 
\label{rimet}
\end{equation}
Observations in the accelerated frame can be interpreted equivalently as observations  of a stationary observer in a gravitational background. Indeed the Rindler metric accounts for  the gravitational field in the near horizon limit of a Schwarzschild black hole (cf.\,\cite{SULE05}) with the black hole horizon at $x=t=0$ and the acceleration given by the inverse of the  Schwarzschild radius. In the non-relativistic limit,  $a\tau \ll 1$, the Rindler metric yields a linear potential
\begin{equation*}U\sim \frac{c^2}{2}\big(g_{00}-1\big)=
\frac{c^2}{2}\big(e^{2a\xi}-1\big)\approx a c^2 \xi\,,\end{equation*} generating  the  parabolic motion of test particles around $t=0$ (cf. Fig.\,\ref{kin}).
\vskip -.3cm
\section{Scalar fields in Rindler spaces}
\begin{figure}
\hspace{-8cm}\includegraphics[width=.55\linewidth]{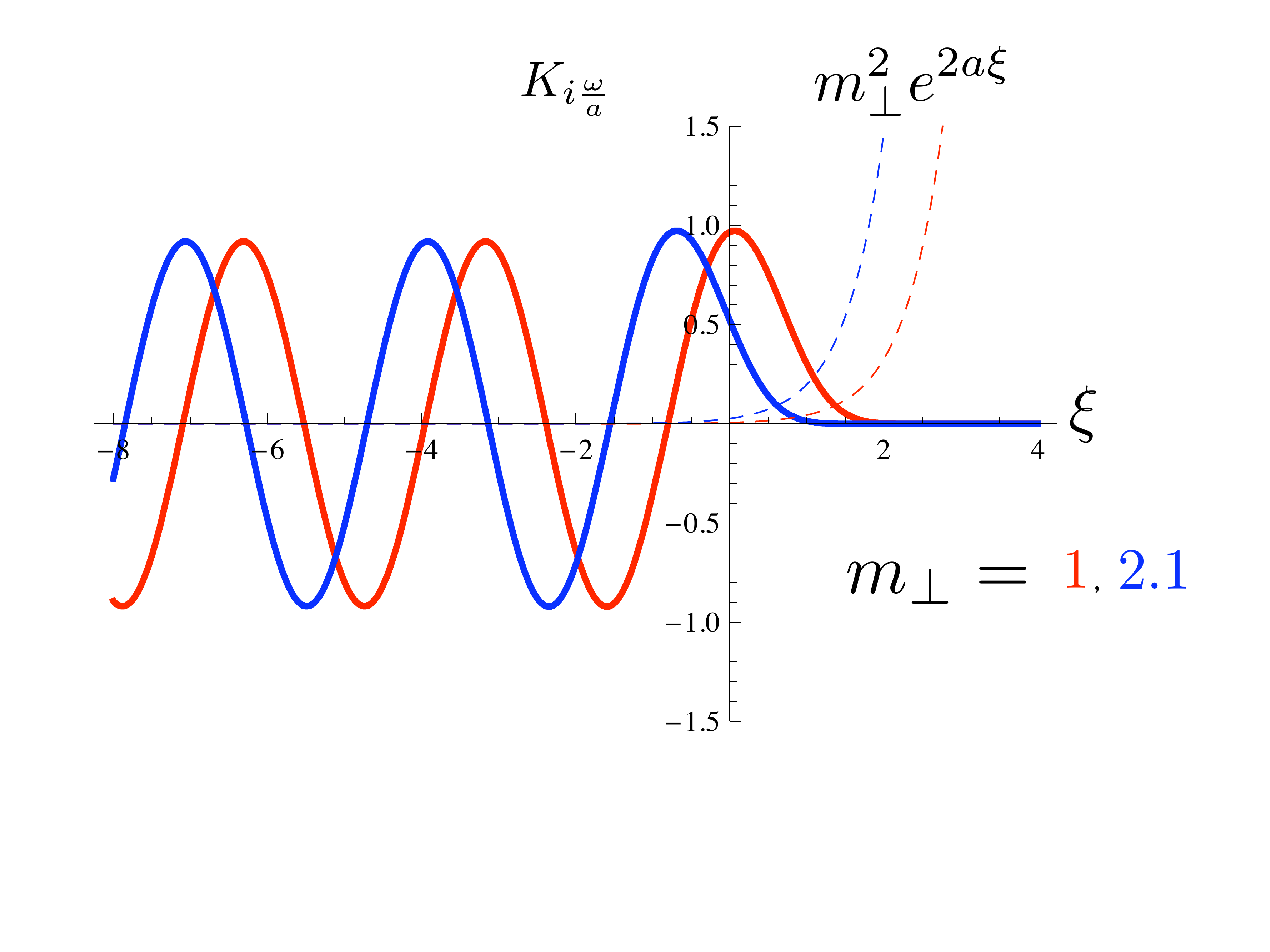}\vskip -6.5cm\hspace{8.3cm}\includegraphics[width=.44\linewidth]{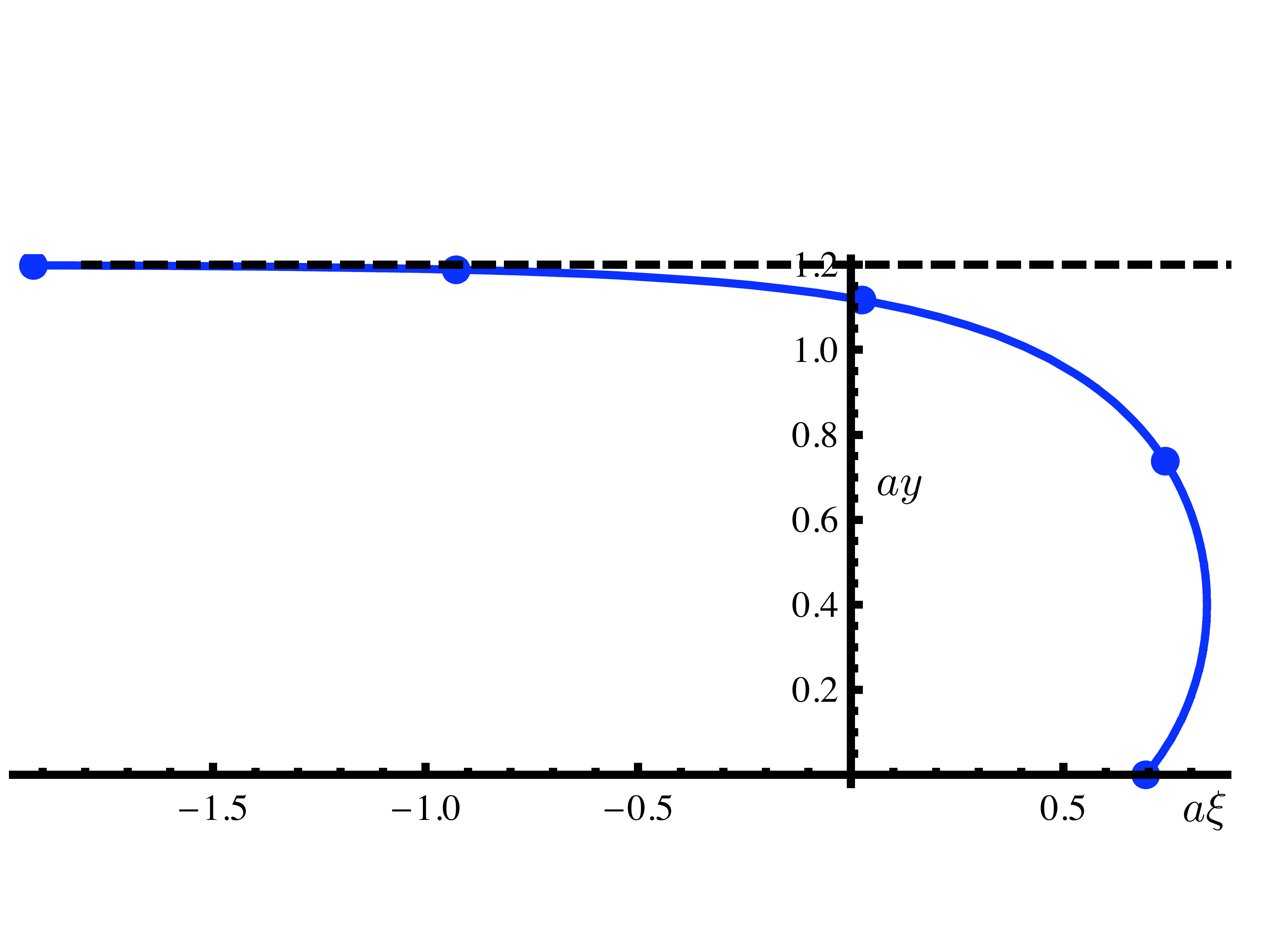}
\vskip -.2cm
\caption{Left: Eigenfunctions of the wave equation (\ref{wxi}) for $\omega/a=2$ and two values of the transverse mass (\ref{trama}) and corresponding potentials of the wave-equation (\ref{wxi}). Right: Trajectory of a particle moving with constant speed in  Minkowski space as seen by a uniformly accelerated observer (solid line) and asymptotic value of the transverse position (dashed line). The points indicate the values $a\tau = 0,1,2,3,4\,.$}
\label{Mcd}
\end{figure}

In preparation for the  discussion of gauge fields I will introduce some of the relevant concepts in the context of  scalar fields in Rindler spaces.  The action of a non-interacting scalar with self-interaction $V(\phi)$ in curved space-time is given by 
\begin{equation*} S = \int \sqrt{|g|} d \tau \; d \xi \; d^{\,d-1}x_{\perp} \big\{ \frac{1}{2} g^{\mu\nu} \partial_{\mu}\phi\partial_{\nu}\phi- V(\phi)  \big \}\,, \end{equation*}
with $g$ denoting the determinant of the metric.
In Rindler space, and keeping only a mass term this expression reduces to 
\begin{equation}S = \frac{1}{2} \int d \tau \; d \xi \; d^{\,d-1}x_{\perp} \big \{ (\partial_{\tau} \phi )^{2}
- (\partial_{\xi} \phi )^{2} - (m^{2} \phi^{2} + (\boldsymbol{\partial}_{\perp} \phi)^{2} ) \;
 e^{2 a \xi} \big \}\,.
\label{scafi}
\end{equation}
Since  $\sqrt{|g|} g^{00}=1$, the expression for the canonical momentum is standard 
\begin{equation*}
\pi(x)=\partial_{\tau} \phi(x)\,,
\end{equation*}
and so is the equal time commutator for $\phi$ and $\pi$.
The fields are expanded in terms of the normal modes of the associated wave equation 
\begin{equation}\phi (\tau, \xi, {\bf x}_{\perp}) \sim \int \frac{d \omega}{\sqrt{2\omega}}\,d^{\,d-1}k_{\perp} (a (\omega, {\bf k}_{\perp}) e^{- i \omega \tau + i {\bf k} _{\perp} {\bf x}_{\perp}}
+ \text{h.c.} )\,
\, K_{i \frac{\omega}{a}} \Big(m_{\perp}\,e^{a\xi}\Big)\,, 
\label{scanom}
\end{equation}
with 
\begin{equation}
m_{\perp}^2=(m^2+{\bf k}_{\perp}^2) /a^2\,, 
\label{trama}
\end{equation}
and  the normalization chosen such that the commutator of creation and annihilation operators $a^{(\dagger)} (\omega, {\bf k}_{\perp}) $ is  standard. 
The non-trivial $\xi$-dependence,  given by the McDonald functions $ K_{i \frac{\omega}{a}} \Big(m_{\perp}\,e^{a\xi}\Big) $  satisfying the equation
\label{waeq}
\begin{equation}
\big(- \frac{d^{2}}{d \xi^{2}} + a^2 m^{2}_{\perp} \; e^{2 a \xi}-  \omega^{2}\Big)  K_{i \frac{\omega}{a}} \big(m_{\perp}\,e^{a\xi}\big)=0 \,     \label{wxi}
\end{equation}
is illustrated in Fig.\,\ref{Mcd}.
The repulsive exponential potential prevents propagation of the wave for positive $\xi$. This repulsion accounts for the fact that a particle moving  with arbitrary constant speed in Minkowski space is seen by the accelerated observer to approach  $\xi=-\infty$ and  the speed of light  for large times $\tau$. This is shown in Fig.\,\ref{Mcd} for a particle starting at $\tau=0$ at $ax_0=2$ with velocity $v^x=0.5\,,\; v^y=0.3$. At  $a\tau = 0.55$, the $x$-component of the  the velocity of accelerated observer and particle coincide and   the $\xi$-component of the  velocity of the particle vanishes. In the accelerated frame, the transverse velocity of the particle vanishes exponentially for large times ($\sim \exp\{-2a\tau\}$) as a result of the forever increasing time dilation induced by the acceleration. Around $\tau=3$,  the particle has essentially reached its asymptotic transverse position $a \tilde{y}^y_{\infty} = ax_0v^y/(1-v^x)$ in the accelerated frame.  
\subsection{Scale invariant Hamiltonian}\label{SiH}
The Hamiltonian of the scalar field in Rindler space is given by
\begin{equation}
H = \int d^{\,d-1}k_{\perp} \int^{\infty}_{0} d \omega\, \omega  a^{\dagger} (\omega, {\bf k}_{\perp}) a (\omega, {\bf k}_{\perp})\,.
\label{hamsc}
\end{equation}
Unlike in Minkowski space, the energy $\omega$ of Rindler particles is independent of the transverse momentum ${\bf k}_{\perp}$ and mass though the solutions  of the wave equation (\ref{wxi}) depend non-trivially on these quantities as is illustrated in  Fig.\,\ref{Mcd}  for two eigenfunctions with the same energy. The kinematical origin of the  degeneracy of the energy eigenvalues  with respect to the transverse momenta is the time dilation in the accelerated frame leading to asymptotically vanishing transverse velocities. The degeneracy can also be interpreted as a consequence of a symmetry. For a massless scalar field, the invariance of the action (\ref{scafi}) under the scale transformation  
\begin{equation}\tau^\prime =\tau\,,\quad e^{a\xi^{\prime}} = e^{a\xi_0} e^{a\xi}\,,\quad  {\bf x}^{\prime} _{\perp}= e^{a \xi_{0}} {\bf x}_{\perp}\,,\label{sctr}\end{equation}
\begin{equation*} \quad\phi (\tau, \xi, {\bf x}_{\perp}) \longrightarrow e^{(d-1)a \xi_{0}/2}\, \phi (\tau, \xi^{\prime},  {\bf x}^{\prime}_{\perp})\,,
\end{equation*}
is manifest. 
The  generator of this transformation 
\begin{equation*}Q = \int d \xi d^{d-1} x_{\perp} \pi (\xi, {\bf x}_{\perp})  \left \{ \partial_{\xi} + a \left (\frac{d-1}{2} + 
{\bf x}_{\perp} \boldsymbol{\partial}_{\perp} \right )  \right \} \phi \end{equation*}
commutes with the Hamiltonian but not with the transverse component of the momentum operator 
\begin{equation}\big[ H_{m=0}, Q \big] = 0, \quad \big[P_i,H_{m=0}\big]=0\,, \quad\big [ P_{i}, Q \big ]   =   - i a P_{i}  \,,\quad i\neq 1\,. 
\label{opal}
\end{equation}
Together with the rotational symmetry in the transverse space these commutators  imply the degeneracy of the Hamiltonian (\ref{hamsc}).
In the presence of a mass term, the generator of the symmetry transformations reads
\begin{eqnarray*}&&Q = \int d \xi d^{d-1} x_{\perp} \pi (\xi, {\bf x}_{\perp})
\Bigg \{\partial_{\xi} + a \Big(\frac{d-1}{2} + 
{\bf x}_{\perp} \boldsymbol{\partial}_{\perp} \Big)   \phi 
+a\,m^{2} \Big (\frac{d-3}{2} + {\bf x}_{\perp} \boldsymbol{\partial}_{\perp} \Big)\\ &&\int 
d^{d-1} y_{\perp} g(|{\bf x}_{\perp}-{\bf y}_{\perp}|)\phi (\tau, \xi, {\bf y}_{\perp})\Bigg\}\quad\text{with} \quad\Delta_{\perp}\, g({\bf x}_{\perp}) = -\delta({\bf x}_{\perp})\,. \end{eqnarray*}
It is remarkable that  the dimensionful Rindler space Hamiltonian is invariant under scale transformations and  that  this invariance persists in the presence of a mass term. 
\subsection{Unruh heat bath}
Here I will discuss the dynamics of scalar fields in Rindler space  in terms of the dynamics  in  Minkowski space observed in a uniformly accelerated frame. The starting point for  establishing  the relation between the scalar field theory in inertial and accelerated frames is the identity of the fields in accelerated and inertial frames (Rindler wedge)  
\begin{equation*}\phi (\tau,\xi,{\bf x}_{\perp}) = \tilde{\phi} (t,{\bf x}) \; \Big |_{t,{\bf x} = t,{\bf x} (\tau, \xi)} \,.
\label{idesca}
\end{equation*}
Projection of this equation onto the Rindler space normal modes (\ref{scanom})
yields the following relation (Bogoliubov transformation)  between the creation and annihilation operators in the two frames
\begin{equation}a (\Omega, {{\bf k}}_{\perp}) = \frac{1}{\sqrt{a \sinh \pi \frac{\Omega}{a}}} \int^{\infty}_{- \infty}
\frac{dk}{\sqrt{4 \pi  \omega_k}}  e^{i \frac{\Omega}{a} \beta_{{\bf k}}} \left [
e^{\frac{\pi \Omega}{2 a}} \tilde{a} (k, {\bf k}_{\perp})  + e^{- \frac{\pi \Omega}{2 a} } \tilde{a}^{\dagger}
(k, -{{\bf k}}_{\perp}) \right ]\,.
\label{cran}
\end{equation}
Observations in the accelerated frame are performed  in the Minkowski ($|0_M\rangle$) rather than the Rindler space vacuum (barring a local cooling in the observers rocket). A fundamental quantity is the number of particles measured in the accelerated frame which with the  help of (\ref{cran}) is found to be 
\begin{equation}\langle 0_M | a^{\dagger} (\Omega, {\bf k}_{\perp})  a (\Omega^{\prime}, {\bf k}_{\perp}^{\prime}) | 0_M \rangle
=\frac{1}{e^{2 \pi \frac{\Omega}{a}} - 1} \delta (\Omega - \Omega^{\prime})\delta({\bf k}_{\perp}-{\bf k}^{\prime}_{\perp}) \,.
\label{thdi}
\end{equation}
In the accelerated frame, a thermal distribution of (Rindler) particles is observed. For a detailed discussion of the ''Unruh effect'' cf. \cite{UNWA84,CRHM07}\,. I will discuss the Unruh radiation in the context of the electromagnetic field. 
\section{Gauge fields in Rindler spaces}
Starting point of the canonical quantization of the electromagnetic field  is the Rindler space action in the Weyl gauge  $A_0=0$ 
\begin{eqnarray}
S[A] &=&\frac{1}{2} \int d \tau d \xi d^{d-1} x_{\perp} \Big \{ e^{- 2 a \xi} ( \partial_{0} A_{1})^{2} + \sum^{d}_{I = 2} \Big [ ( \partial_{0} A_{I})^{2} \nonumber\\ &-& 
(\partial_{1} A_{I} - \partial_{I} A_{1} )^{2} \Big ] - e^{2 a \xi} \sum^{d}_{J > I = 2}
(\partial_{I} A_{J} - \partial_{J} A_{I})^{2} \Big \}\,.
\label{actem}
\end{eqnarray}
In the Weyl gauge, the  quantization of gauge fields  in Rindler space is standard
\begin{equation*}\Pi_i=\partial_0\,A^i,\quad [\Pi^{i}(\tau,\xi,{\bf x}_{\perp}), A_{j}(\tau,\xi^{\prime},{\bf x}_{\perp}^{\prime})] 
= \frac{1}{i}\delta_{i j } \delta(\xi-\xi^{\prime})\delta({\bf x}_{\perp} - {\bf x}_{\perp}^{\prime})\,. \end{equation*}
The Gau\ss\ law is implemented as a constraint on the space of physical states
\begin{equation*}\partial_i \,\Pi^i\,|\psi\rangle = 0 \,, \end{equation*}
and the Hamiltonian reads
\begin{equation}H = \frac{1}{2} \int  d \xi d^{\,d-1}x_{\perp} \Big\{ e^{2a\xi} \Pi^{1\,2}+ \sum_{I=2}^d \Pi^{I\,2} 
+ \sum_{I=2}^d \big( F_{1I}^2 + e^{2a\xi}\sum_{J>I}^d  F_{IJ}^2 \big)\Big\}\,.
\label{hammex}
\end{equation}
Following  the method developed in section \ref{SiH}  one easily derives the degeneracy of the energy eigenstates without explicit construction of the normal modes of the system. Under the scale transformations (cf. Eq.\,(\ref{sctr})) 
\begin{eqnarray}A_{1} (\tau, \xi, {\bf x}_{\perp}) &\to& e^{(d-3)a \xi_{0}/2} A_{1} (\tau, \xi^{\prime}, {\bf x}_{\perp}^{\prime})\,, \quad\nonumber\\
A_{I} (\tau, \xi, {\bf x}_{\perp}) &\to&  e^{(d-1) a \xi_{0}/2} A_{I} (\tau, \xi^{\prime}, {\bf x}_{\perp}^{\prime})\,, 
\label{stgf}
\end{eqnarray}
the action (\ref{actem}) remains invariant and the generator of the symmetry transformations is given by 
\begin{equation}Q_A =\int d \xi d x_{\perp} \Bigg\{ \sum^{d}_{i = 1} \Pi^{i} \Big[\partial_{\xi} + a\big(\frac{d-1}{2}+{\bf x}_{\perp}\boldsymbol{\partial}_{\perp}\big)\Big]A_{i} -a \,\Pi^1 A_1\Bigg\}\,.
\label{QA} 
\end{equation}
The generator $Q_A$, the Hamiltonian $H_A$ and the transverse momentum operator of the electromagnetic field  satisfy the operator algebra (\ref{opal}) with the same consequences for the spectrum as for scalar fields.
\subsection{QED and QCD in Rindler space time}
The transformation property of the covariant derivative
\begin{eqnarray}\partial_1-i e A_1(\tau,\xi,{\bf x}_{\perp}) &\to& \partial_1^{\prime} -i \hat{e}(\xi_0)\, A_1 (\tau, \xi^{\prime}, {\bf x}_{\perp}^{\prime})\,,\quad \hat{e}(\xi_0) = e^{(d-3)a\xi_0/2} e \,,\nonumber\\
\partial_I-i e A_I(\tau,\xi,{\bf x}_{\perp}) &\to& e^{a\xi_0} \big(\partial_I^{\prime} -i\hat{e}(\xi_0)\, A_I (\tau, \xi^{\prime}, {\bf x}_{\perp}^{\prime})\big)\,,
\label{covdev}
\end{eqnarray}
is essential for the  symmetry analysis of interacting gauge theories. The Rindler space action of the electromagnetic field  coupled to massless fermions  (cf.\cite{BRWH57,NAKA90})
\begin{eqnarray*}&&S[A,\psi,e] = S[A]+\int d \tau d \xi d^{d-1} x_{\perp}\\ && \bar{\psi}\, i\Big \{\gamma^{0}\big(\partial_{0} -ieA_{0}\big) +\gamma^{1}\big(\partial_{1} +\frac{a}{2}-ieA_{1}\big) +e^{a\xi}\gamma^I \big(\partial_I -ie A_I\big)\Big\}\psi \,,\end{eqnarray*}
transforms under the combined transformation of gauge (\ref{stgf}) and fermion fields
\begin{equation}\psi(\tau,\xi, {\bf x}_{\perp}) \to e^{(d-1)a \xi_{0}/2} \psi (\tau, \xi^{\prime}, {\bf x}^{\prime}_{\perp})\,
\label{fetr}
\end{equation}
as 
$$ S[A,\psi,e] \to S[A,\psi,\hat{e}(\xi_0)]\,,$$
with generator (cf. (\ref{QA}))
$$Q = Q_A-i\int d \xi d x_{\perp} \psi^{\dagger}\Big[\partial_{\xi} + a\big(\frac{d-1}{2}+{\bf x}_{\perp}\boldsymbol{\partial}_{\perp}\big)\Big]\psi\,, $$
and with the following change in the coupling constant 
\begin{equation}\hat{e}(\xi_0) = e^{(d-3)a\xi_0/2} e\,.
\label{ruco}\end{equation}
This result applies also to Yang-Mills theories coupled to massless quarks. The transformation of gauge fields (\ref{stgf}), of covariant derivatives (\ref{covdev}) and  of fermion fields  (\ref{fetr}) have to be applied to each color component and the coupling constant $e$ in (\ref{ruco}) replaced by the Yang Mills coupling constant $g_{YM}$ with the result
\begin{equation*}S[A,\psi,g_{YM}] \to S[A,\psi,\hat{g}_{YM}(\xi_0)]\,.\end{equation*}
In 3+1 dimensional Rindler space time (d=3) the combined transformation leaves the (tree level) Hamiltonians invariant
\begin{equation*}[H_{QED}, Q ] =0\,,\quad [H_{QCD}, Q ] =0\,,\end{equation*}
implying degeneracies in the spectra.
\subsection{Electromagnetic fields in accelerated frames}
In order to study detailed properties of electromagnetic fields in Rindler space, the Gau\ss\ law constraint has to be resolved and the normal modes have to be constructed \cite{LEOY08}.  The resolution of the Gau\ss\ law constraint is most efficiently achieved by decomposition of the  gauge fields in transverse and longitudinal fields 
\begin{equation}\hat{A}^{i} = A^{i} + \partial^{i}\,\frac{1}{\Delta} \;
\partial_{j} A^{j}\,,\quad \Delta = - \partial_{i} \partial^{i}= \partial_{\xi}e^{-2a\xi}\partial_{\xi}+\boldsymbol{\partial}_{\perp}^2\,.
\label{trvlap}
\end{equation}
The normal mode decomposition of the transverse field operators $\hat{A}^{i}$ is carried out in terms of the d-1 annihilation and creation operators
$a_i^{(\dagger)}(\omega,{\bf k}_{\perp})$. The normal modes are given by the McDonald functions (\ref{wxi}) and powers of their argument. In order to establish the relation between measurements in inertial and accelerated frames we again have to identify appropriate field operators in the two systems (cf. Eq.\,(\ref{idesca})). Here the  transverse gauge field $\hat{A}_i(x)$ in the accelerated frame $x = \{\tau,\xi,{\bf x}_{\perp}\}$ has to be identified with the transformed inertial frame (transverse) gauge field $\hat{\tilde{A}}_i(\tilde{x})$ in the Rindler wedge $\tilde{x} = \{t,x^1,{\bf x}_{\perp}\}=\tilde{x}(\tau,\xi,{\bf x}_{\perp})$ 
\begin{equation}\hat{A}_i(\tau,\xi,{\bf x}_{\perp})=\frac{\partial \tilde{x}^k}{\partial x^{i}}\hat{\tilde{A}}_k(\tilde{x}(\tau,\xi,{\bf x}_{\perp}))-\frac{\partial }{\partial x^{i}} \int^{\tau} d\tau^{\prime}\,\frac{\partial \tilde{x}^k}{\partial \tau^{\prime}}\hat{\tilde{A}}_k (\tilde{x}(\tau^{\prime},\xi,{\bf x}_{\perp}) ) \,.
\label{gftr}
\end{equation}
The first term on the right hand side accounts for the coordinate transformation of a vector field, the second  represents the necessary gauge transformation to the Weyl gauge. 
Projection on the normal modes of the components of the transverse gauge fields  yields as above (\ref{cran}) the expression of the Rindler space creation and annihilation operators in terms of the corresponding Minkowski space operators  
\begin{eqnarray}a_i (\Omega, {{\bf k}}_{\perp}) &=& \frac{1}{\sqrt{a \sinh \pi \frac{\Omega}{a}}} \int^{\infty}_{- \infty}
\frac{dk}{\sqrt{4 \pi  \omega_k}}  e^{i \frac{\Omega}{a} \beta_{{\bf k}}}\sum_{j=1}^{d-1}\Big[e^{\frac{\pi \Omega}{2 a}} R_{ij}(k,{\bf k}_{\perp}) 
\tilde{a}_j (k, {\bf k}_{\perp}) \nonumber\\  &+& e^{- \frac{\pi \Omega}{2 a} }R_{ij}(k,-{\bf k}_{\perp})  \tilde{a}_j^{\dagger}
(k, -{{\bf k}}_{\perp}) \Big]\,,
\label{crab}\end{eqnarray}
where the matrix $R$ describes the mixing of the components of the gauge fields  under the coordinate and gauge transformations of Eq.\,(\ref{gftr}).
\subsection{The electromagnetic Unruh heat bath}
\begin{figure}
\begin{center}
\includegraphics[width=0.45\linewidth]{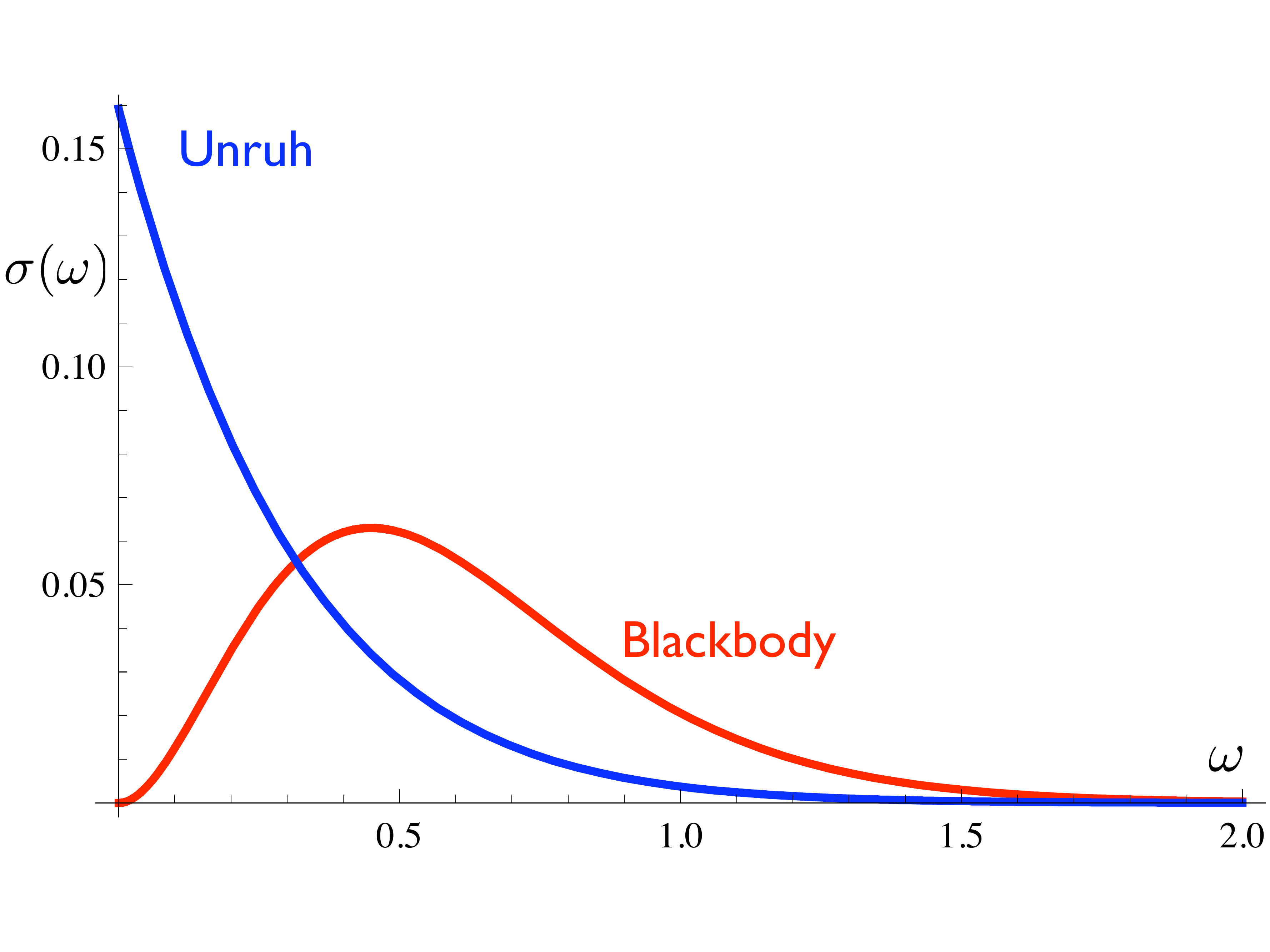}
\caption{The integrand of Eq.\,(\ref{ends}) in comparison with the corresponding integrand of the energy density in the blackbody radiation in 3 space dimensions  as a function of $\omega$ in units of the acceleration ($a$) and the temperature ($T/2\pi$) respectively}
\label{plan1}
\end{center}
\end{figure}

The relation\,(\ref{crab}) of the creation and annihilation operators in inertial and accelerated frames implies the following expression for the number of  Rindler photons in the Minkowski vacuum  (cf. Eq.\,(\ref{thdi})) 
\begin{equation*}\langle 0_M | a^{\dagger}_{i} (\omega, {\bf k}_{\perp}) a_{j} (\omega^{\prime}, {\bf k}_{\perp}^{\prime}) | 0_M \rangle
=\frac{1}{e^{2 \pi \frac{\omega}{a}} - 1} \delta_{i j}\,\delta (\omega - \omega^{\prime})\delta({\bf k}_{\perp}-{\bf k}^{\prime}_{\perp})\,. \end{equation*}
Although of the same structure as the corresponding expression for the number of Minkowski space photons at finite temperature, the different dispersion law of Rindler photons 
$\frac{\partial\, \omega}{\,\partial\, {\bf k}_{\perp}}=0$ 
gives rise to significant changes in  Unruh as compared to  blackbody radiation. I illustrate this by a discussion of the energy density $ {\cal H}$ (cf. Eq.\,(\ref{hammex})) of the Unruh radiation.  As in finite temperature field theory,  divergences in  $ {\cal H}$ are avoided by normal ordering with respect to the (Rindler) ground state and the following result is obtained
\begin{equation}\langle 0_{M} |: {\cal H} (\xi, {\bf x}_{\perp}) :_R| 0_{M} \rangle = N_{d} a^{d+1} e^{(1-d)a \xi} \int_{0}^{\infty}  \frac{1}{a} d\omega\,\sigma\big(\frac{\omega}{a}\big)\,,\label{ends}\end{equation}
\begin{equation*}\sigma(\kappa) =  \frac{\kappa^2 + \frac{1}{4} (d-1)^{2} }{e^{2 \pi\kappa}+(-1)^{d}}\left \{ 
\begin{array}{l} \;\kappa \; P^{o}_{\frac{d-3}{2}}
\left ( \kappa^2 \right )  \\[2ex]
 \,\quad P^{e}_{\frac{d-2}{2}} \; \left ( \kappa^2 \right ) 
\end{array}\right .    \,, \end{equation*}
where $ P^{o}_{\frac{d-3}{2}}$ and $ P^{e}_{\frac{d-2}{2}}$ denote everywhere non-zero polynomials  for odd and even space dimensions respectively with the degree given by the index and  with asymptotics 
\begin{equation*}\kappa \; P^{o}_{\frac{d-3}{2}}(\kappa^2) ,\quad    P^{e}_{\frac{d-2}{2}} \; (\kappa^2) \xrightarrow [\kappa\to \infty]{} (\kappa)^{d-2} \,.\end{equation*}
After redefining the energy density  with respect to the measure $\sqrt{|g|} d\xi dx_{\perp}$ instead of the flat measure (\ref{hammex}),  
${\cal H}^{\prime} (\xi, {\bf x}_{\perp})= e^{-2a\xi}{\cal H} (\xi, {\bf x}_{\perp})\,,$ Eq.\,(\ref{ends}) is  rewritten as  
$$\langle 0_{M} |: {\cal H}^{\prime} (\xi, {\bf x}_{\perp}) :_R| 0_{M} \rangle  \sim  \big(a e^{-a \xi}\big)^{d+1} \sim T_{\xi}^{d+1} \,,$$
with  $T_\xi$ satisfying   Tolman's law of relativistic thermodynamics \cite{TOLM87}
$$T_{\xi}\sqrt{g_{00}} = \text{const}\,.$$
As is illustrated in Fig.\,\ref{plan1}, the integrand in Eq.\,(\ref{ends}) exhibits significant differences in the frequency distributions of Unruh and blackbody radiation respectively. The two integrands agree in their high frequency behavior ($\sim \omega ^d e^{-\omega}$); at threshold the integrand of the Unruh radiation approaches a constant while that of the blackbody radiation vanishes like $\sim \omega^{d-1}$.
\\The appearance of a photon condensate is another consequence of the different dispersion laws of Rindler and Minkowski photons. Unlike in Minkowski space at finite temperature, in  Rindler space the photon condensate is in general different from zero
\begin{eqnarray*}\langle 0_{M} |: {\cal H}_{E} - {\cal H}_{B}:_R | 0_{M} \rangle &=& N_{d} d^{d} \; 
e^{(1-d)a \xi} \frac{(d-1)(3-d)}{2}\\
&\cdot&\int_0^\infty d \omega \frac{1}{e^{2 \pi \frac{\omega}{a}}+(-1)^{d}}
\left \{ 
\begin{array}{l}   \, \frac{\omega}{a} \; 
P^{o}_{\frac{d-3}{2}} \; \left ( \frac{\omega^{2}}{a^{2}} \right ) \\[2ex]
 P^{e}_{\frac{d-2}{2}} \; 
\left ( \frac{\omega^{2}}{a^{2}} \right )
\end{array}  \right. \,.\end{eqnarray*}
It vanishes for $d=3$ and is dominated for higher dimensions by the magnetic field contribution.
\subsection{Interaction energy of static charges in accelerated frames}
In the last application I will discuss the interaction energy of two uniformly accelerated  static charges measured by a co-accelerated observer.     
The Coulomb energy of oppositely charged pointlike sources in Rindler space is given by
\begin{equation*}V_{C}=e^2 
D (\xi_1, {\bf x}_{\perp}, \xi_2, {\bf 0}_{\perp})\,, \quad D({\bf x},{\bf x}^{\prime}) = \langle {\bf x} |\frac{1}{\Delta}| {\bf x}^{\prime}\rangle \,,\end{equation*}
with the Laplace operator $\Delta$ defined in (\ref{trvlap}). The static propagator can be evaluated analytically in terms of Legendre functions of the second kind 
\begin{equation}\frac{V_C}{a} = -\frac{e^{- i \frac{\pi}{2} (d-2)}}{2^{3/2} \pi^{d/2}} \; 
\tilde{e}^{2}\; (u^{2}-1)^{- (d-2)/4}  \; Q^{\frac{d}{2}-1}_{1/2} (u)\,.
\label{VCO}\end{equation}
The quantity $u(\xi_1, {\bf x}_{\perp\,1}, \xi_2, {\bf x}_{\perp\,2})$ is given  by the geodesic distance  $d_g$ in units determined by the average of the coordinates $\xi_1,\,\xi_2$ of the 2 charges
\begin{equation*}
u = 1+\frac{1}{2}s^{2}\,, \quad s^{2} = \frac{d^2_g }{a^{-2} e^{a(\xi_1+\xi_2)}}\, ,\quad d^2_g=a^{-2}(e^{a\xi_1} - e^{a\xi_2})^{2}+ ({\bf x}_{\perp\,1}^{2}-{\bf x}_{\perp\,2}^{2})^2\,.\end{equation*}
\begin{figure}
\hspace{-8.4cm}
\includegraphics[width=.58\linewidth]{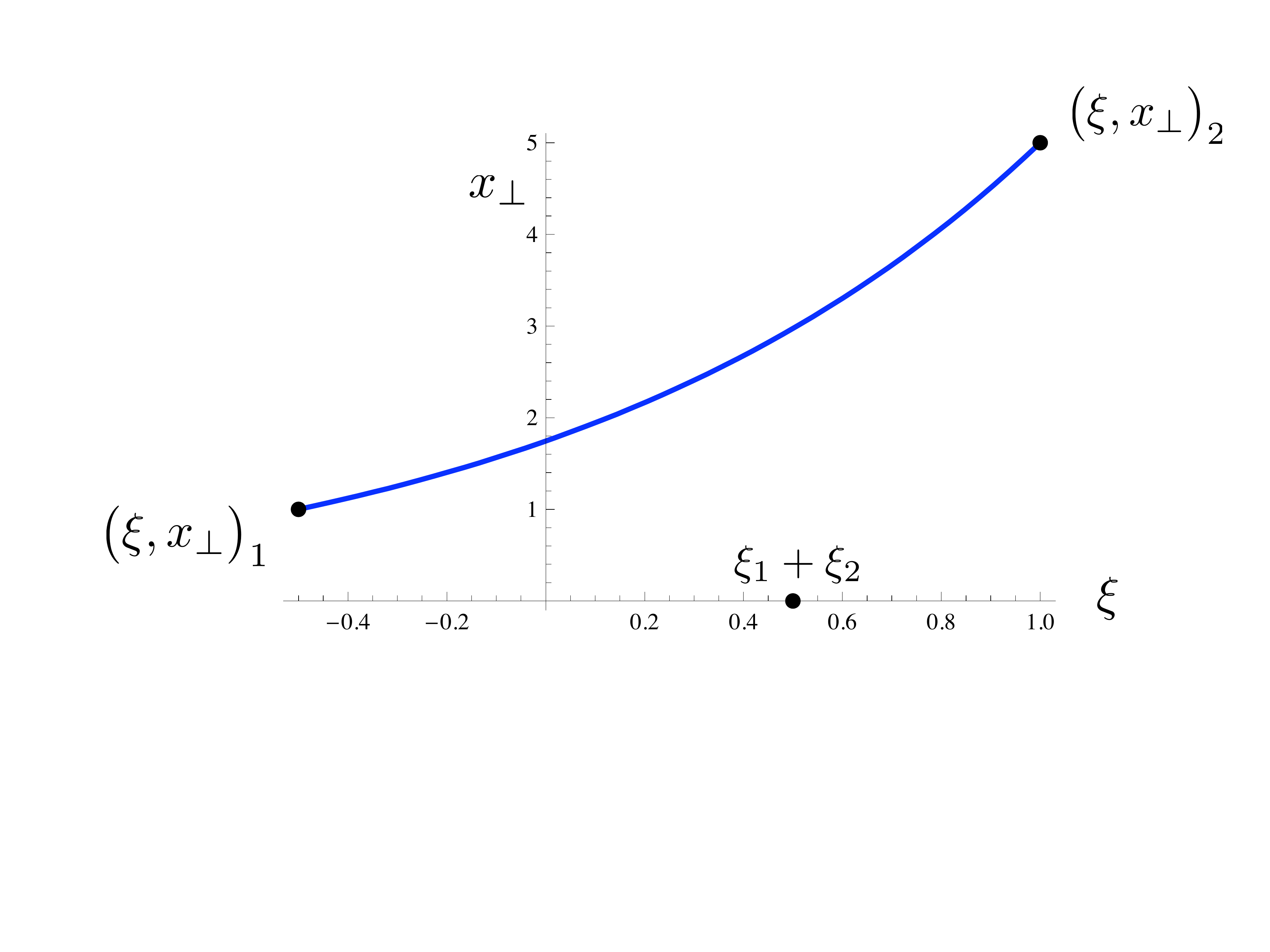}
\vskip-6.60cm \hspace{8.cm}
\includegraphics[width=.48\linewidth]{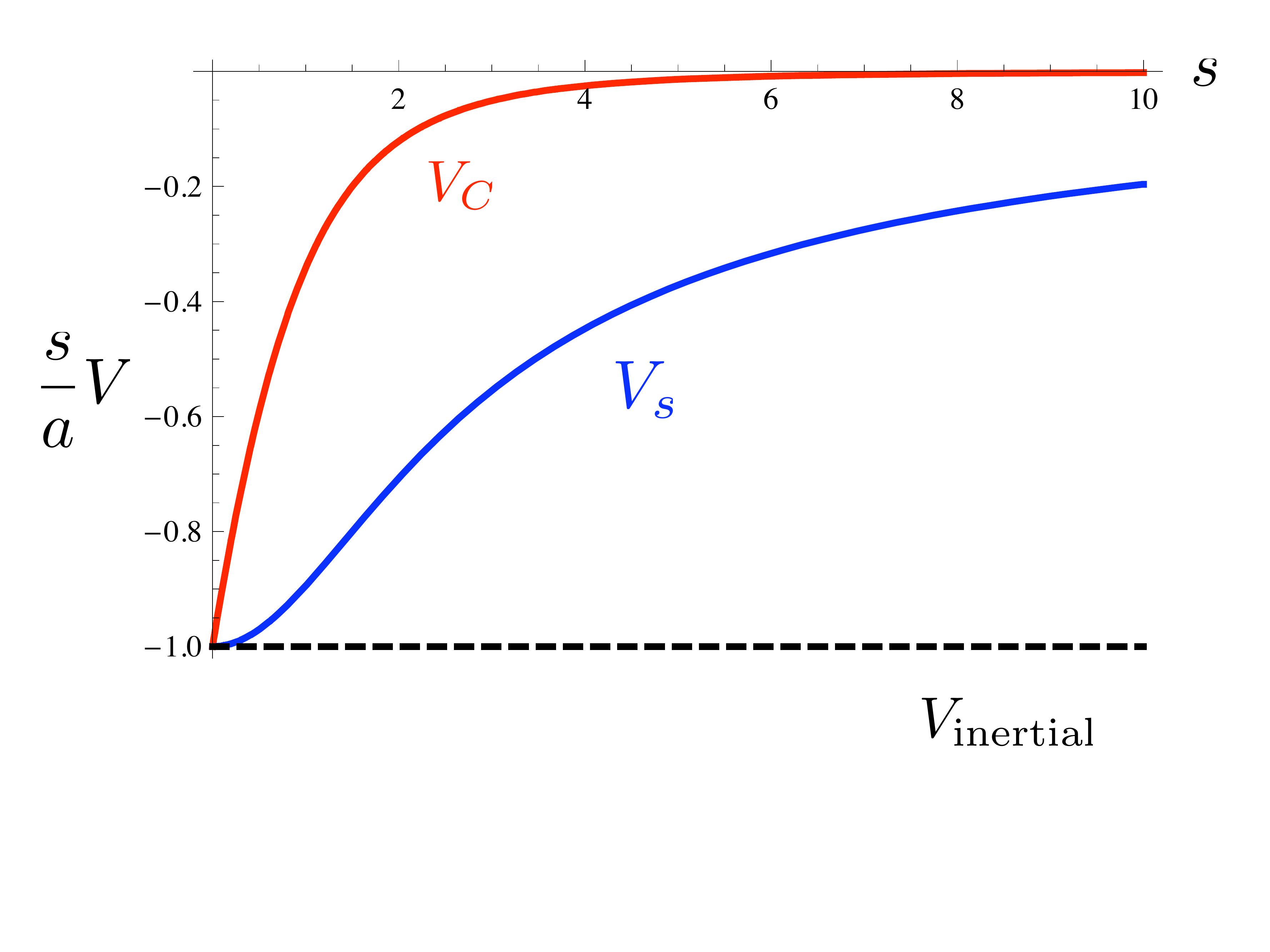}
\vskip -1cm
\caption{Left: The geodesic connecting two charges at equal time. Right: Interaction energy of  static  vector ($V_C$) and scalar ($V_s$) charges in   3+1 dimensional Rindler  space multiplied with their distance in comparison to the same quantities in  Minkowski space}
\label{vcvs}  
\end{figure}
The left side of Fig.\,\ref{vcvs} shows the geodesic connecting the two charges. 
The strength of the interaction $\tilde{e}^2$ 
also ''runs''  with the average $\xi$ coordinate.  Its definition  
$$ \tilde{e}^{2} = \frac{e^2 }{(a^{-2} e^{a(\xi_1+\xi_2)})^{(d-3)/2}} = \Big(\frac{a}{\sqrt{2}}\Big)^{d-3} \hat{e}^2\big(-\frac{1}{2}(\xi_1+\xi_2)\big) \,,$$ 
is in accordance with the definition (\ref{ruco})  of $\hat{e}$  determined from the scaling properties of the action of the Maxwell field. The running of $\tilde{e}$ with the $\xi$ coordinates of the sources is  reminiscent of the  Ads/CFT duality where the  scale is set by the coordinate transverse to the 4-dimensional Minkowski like space \cite{POST01}.
The interaction energy displays power law behavior for large and small distances 
\begin{equation*}\frac{V_C}{a} \xrightarrow [s\to 0]{}   - \frac{\tilde{e}^2}{s^{d-2}} \,, \quad\frac{V_C}{a} \xrightarrow [s\to \infty]{}   - \frac{\tilde{e}^2}{s^{d+1}}\,. \end{equation*}
At small distances it agrees with the inertial frame result and is suppressed by three powers at large distances. 
In three space dimensions (as for any odd $d$), the interaction energy (\ref{VCO}) can be expressed by elementary functions 
\begin{equation} \frac{V_C}{a}= \frac{e^2}{4\pi}   \Big( 1 - \frac{1+\frac{1}{2}s^2}{\sqrt{s^{2} +\frac{1}{4}s^4}}\Big)\,.
\label{VC3}
\end{equation}
In Fig.\,\ref{vcvs} this interaction energy in an accelerated frame is compared with  the inertial frame result and with the interaction energy $V_s$ of sources coupled to a massless scalar field. Common to the three interaction energies is the familiar ''$1/r$'' behavior at small distances while at large distances scalar and vector interaction energies differ from each other and both from the inertial frame result.  In Rindler space, the scalar propagator is the inverse of (cf.(\ref{trvlap})) 
\begin{equation*}\Delta_s = - e^{2a\xi}\partial^{i} \partial_{i}= \partial_{\xi}^2+e^{2a\xi}\boldsymbol{\partial}_{\perp}^2\,, \end{equation*}
and, independent of the dimension, the following ratio of interaction energies is obtained  
\begin{equation*}\frac{V_{s}}{V_C}=e^{-a(\xi_1+\xi_2)} \Big(1+ \frac{1}{2}s^2+s\sqrt{1+\frac{1}{4}s^2}\,\Big)\,.\end{equation*}
\section{Conclusions}
I conclude with a few remarks, concerning  Yang-Mills fields and QCD as seen by an accelerated observer.  Due to its  kinematical origin, the characteristic Rindler space degeneracy  must show up also in the spectrum of interacting theories.  Whatever the elementary excitations, their energy measured by an accelerated observer will neither depend on the transverse momentum, nor on the mass of e.g.  glueballs or mesons.  Furthermore the analogy with the finite temperature Yang-Mills theory or QCD leads one to expect that either,  as a function of the acceleration,  these systems  exhibit a phase transition  or,  because of the high degeneracy, do not  exhibit at all a confined or a chirally broken phase. In line with such an expectation is the observed weakening of the electrostatic  interaction  of charges at large distances.
\section*{Acknowledgment}
I thank K. Yazaki for discussions and for a careful reading of the manuscript.

\end{document}